# Electronic Properties of MoS$_2$/MX$_2$/MoS$_2$ Trilayer Heterostructures: A First Principle Study

*Kanak Datta and Quazi D. M. Khosru
Department of Electrical and Electronic Engineering
Bangladesh University of Engineering and Technology
Dhaka-1000, Bangladesh.
*Email: kanakeee08@gmail.com

## Abstract

In this work, we have presented a first principle simulation study on the electronic properties of MoS$_2$/MX$_2$/MoS$_2$ (M=Mo or W; X=S or Se) trilayer heterostrcuture. We have investigated the effect of stacking configuration, bi-axial compressive and tensile strain on the electronic properties of the trilayer heterostructures. In our study, it is found that, under relaxed condition all the trilayer heterostructures at different stacking configurations show semiconducting nature. The nature of the bandgap however depends on the inserted TMDC monolayer between the top and bottom MoS$_2$ layers and their stacking configurations. Like bilayer heterostructures, trilayer structures also show semiconducting to metal transition under the application of tensile strain. With increased tensile strain the conduction band minima shifts to K point in the brillouin zone and lowering of electron effective mass at conduction band minima is observed. The study on the projected density of states reveal that, the conduction band minima is mostly contributed by the MoS$_2$ layers and states at the valance band maxima are contributed by the middle TMDC monolayer.

## Introduction

Layered Transition Metal Dichalcogenides (TMDC) have become a focus of intense research and investigation in recent times. Existence of intrinsic bandgap, possible scaling and stability of electronic properties down to atomic level thickness, dangling free bond at the interface make these materials suitable for next generation ultra-scaled transistor channel applications (1). For example, monolayer MoS$_2$ shows an intrinsic bandgap of about 1.8 eV and a mobility of about 200 cm$^2$/v/s (2). Transistors based on monolayer and few layers of MoS$_2$ have been experimentally demonstrated in recent years with improved output characteristics (2) (3) (4). Apart from experimental investigation, theoretical investigation from first principle studies on these materials also reveals interesting properties (5) (6). First principle studies on monolayer MoS$_2$ reveals transition from direct to indirect bandgap nature, deformation in the bandstructure, gradually reducing effect mass at the conduction band minima when biaxial tensile strain is applied (7) (8). Besides monolayers, bilayer heterostructures and heterojunctions of these layered materials have drawn interest in recent years (9) (10). Theoretical investigation on the electronic properties of bilayer TMDC materials can be found in recent literature (11) (12). Theoretical investigation on these bilayer heterostructures gives a new window in tuning electronic and optoelectronic properties of these materials

(13) (14) (15) (16). For example, although monolayer $MoS_2$ (bandgap 1.8 eV) and $WS_2$ (bandgap 1.9 eV) are both direct bandgap materials, their bilayer heterostructure ($WS_2/MoS_2$) is indirect material with slightly reduced bandgap (1.58 eV) (15). Devices based on heterostructures of TMDC materials have been experimentally investigated and reported in recent literature (17) (18) (19) (20). More recently, theoretical investigation on trilayer heterostructure TMDC materials can be found in literature which shows the application of intercalation as a possible tool to further modulate the electronic properties of TMDC materials (21). In this work, we have performed first principle investigation on electronic properties of trilayer TMDC heterostructures based on $MoS_2$. The simulation has been carried out by open source simulation package Quantum Espresso (22). We have also studied the effect of bi-axial strain on the electronic structure of the trilayer heterostructures. From our first principle simulation, we have reported the values of electron effective mass at different high symmetry points in the brillouin zone.

## Simulation Methodology

The first principle study was carried out using open source simulation package "Quantum Espresso". After geometry optimization of the trilayer structure, ground state energy of the relaxed structure was calculated and band structure calculation was performed. We performed study on three different stacking combinations of $MoS_2/MX_2/MoS_2$ (M=W/Mo; X=S/Se) trilayer heterostructures. A simple schematic representation of the stacking configuration is given in figure 1.

The first principle simulation was carried out using Quantum Espresso, an open source simulation package (22). The simulation methodology started with a geometry optimization of the trilayer lattice structure and then performing a self-consistent field (scf) calculation for bandstructure extraction. For structural optimization, plane wave basis set with cut off energy 80 Ry was used. In our calculation we have considered norm conserving pseudopotentials for all the atoms with Perdew-Burke-Ernzerhof (PBE) exchange correlation functional (23) (24). For structural optimization, we used 18x18x1 Monkhorst-Pack K point meshing for brillouin zone sampling (25). The structural optimization was performed until the force on each atom was less than *0.01eV/Å*. The self-consistent convergence threshold for energy was set at $10^{-9}$ Ry. To avoid interaction between two adjacent images of the unit cell of the trilayer lattice, 30 Å distance was maintained between two adjacent images. *In this investigation, the effect of spin-orbit coupling has not been considered. However, the supplementary material contains bandstructure results with spin orbit coupling.*

After optimization of lattice structure, for better accuracy, scf calculation was performed with brillouin zone sampling of 24x24x1 K points. The output of the scf calculation was used in the extraction of bandstructure. Finally, we plotted the bandstructure along $\Gamma$-M-K-$\Gamma$ direction in the brillouin zone and observed the effect of strain and different lattice stacking configurations.

## Results and Discussion

<u>Study on the Relaxed Trilayer MoS$_2$/MX$_2$/MoS$_2$ Heterostructures:</u>

<u>MoS$_2$/MoSe$_2$/MoS$_2$ Trilayers:</u> As mentioned earlier, first principle simulation was performed on three different stacking configurations of the trilayer heterostructures. As seen from the figure 1(a), for AAA stacking configuration, Se atoms just sit below the S atoms, Mo atom of the middle MoSe$_2$ layers sits in a linear alignment with Mo atoms of top and bottom MoS$_2$ layers. After structural relaxation of the unit cell, the optimized lattice constant was found to be 3.239 Å. The distance between S and Se atoms as shown in figure 1(a) was found to be 3.75 Å. This intermediate distance between two monolayers decreases greatly when the stacking configuration changes to ABA and ACA configurations. From our calculation on geometry relaxation, we obtained interlayer distance of 3.152 Å and 3.1544 Å respectively for ABA and ACA stacking configurations. The lattice constants for these two stacking configurations were found to be about 3.24 Å and 3.238 Å. These values closely matches with the values reported in for trilayer stacking of MoS$_2$ heterostructures (21).

The bandstructure of the trilayer show interesting property as well. Bilayer MoS$_2$/MoSe$_2$ heterostructure is a direct bandgap material with the conduction band minima and valance band maxima occurring at K point in the brillouin zone (15). In our simulation of the trilayer structure, we found similar results. The trilayer heterostructure at stacking configurations AAA was found to be direct bandgap material with bandgap 0.707 eV with the conduction band minima and valance band maxima occurring at the K point in the brillouin zone. However, in case of ABA and ACA configurations, we found the conduction band minima at K point and valance band maxima at Γ point. In these two cases, the valance band maxima at K point was only slightly lower than the valance band maxima at Γ point in the brillouin zone. The extracted bandgaps were found to be 0.729 eV and 0.752 eV for ABA and ACA configurations respectively with both the materials showing indirect bandgap characteristics. The extracted bandstructure of MoS$_2$/MoSe$_2$/MoS$_2$ trilayers is shown in figures 2 (a)-(c).

<u>MoS$_2$/WS$_2$/MoS$_2$ Trilayers:</u> In case of MoS$_2$/WS$_2$/MoS$_2$ trilayer heterostrcuture at AAA stacking, the geometry relaxation of the unit cell lead to interlayer distance of 3.69 Å. In case of AAA stacking, the lattice constant was found to be 3.186 Å. The lattice constant showed almost no change when the stacking was changed to ABA and ACA configurations. For ABA and ACA stacking configurations, the lattice constant of the unit cell was found to be 3.183 Å and 3.186 Å respectively. The interlayer separation for ABA and ACA stacking configurations were found to be approximately 3.12 Å and 3.20 Å which appears to be in good agreement with the results reported in (21).

The calculation on the bandstructure revealed all the stacking configurations to be indirect bandgap materials under relaxed condition. The extracted bandstructure is shown in figures 2 (d)-(f). For all these stacking configurations of MoS$_2$/WS$_2$/MoS$_2$ trilayer, the conduction band minima was found to be occurring at K point and the valance band maxima was found to be occurring at Γ point, therefore making the nature of the bandgap indirect. The calculated bandgaps for AAA, ABA and ACA stackings were 1.30 eV, 1.075 eV and 1.147 eV respectively.

MoS$_2$/WSe$_2$/MoS$_2$ Trilayers: When a WSe$_2$ monolayer was inserted between two MoS$_2$ monolayers, the lattice constant was found to be similar to the case of MoS$_2$/MoSe$_2$/MoS$_2$ trilayer heterostrcuture. The optimized lattice constant for AAA, ABA and ACA stacking configurations were found to be 3.241 Å, 3.237 Å, 3.243 Å. On the other hand, the interlayer separation between the monolayers at AAA, ABA and ACA stacking configurations were found to be 3.781 Å, 3.0848 Å, 3.1788 Å.

In case of bandstructure, all the three stacking configurations were found to be direct bandgap semiconductors under relaxed condition. As seen from the figures 2 (g)-(i), In each case, we found the conduction band minima and valance band maxima to be occurring at K point. The direct bandgap as extracted from bandstructure were found to be 0.45 eV, 0.55 eV and 0.562 eV for AAA, ABA and ACA stacking configurations respectively.

Study on the Projected Density of States of Trilayer Heterotructures:

In order to study the nature of the conduction and valance bands in the bandstructure, we have observed the projected density of states (PDOS) of the d-electrons of the transition metals (Mo/W) and p-electrons of the chalcogen atoms(S/Se). We are reporting the PDOS study on the AAA stacking configuration only.

For the MoS$_2$/MoSe$_2$/MoS$_2$ trilayers, the direct bandgap was found to be 0.707 eV under relaxed condition of the lattice. The study on the projected density of states reveals that the top of the valance band appears to be populated by d-electrons and p-electrons of the middle MoSe$_2$ layers. On the other hand the bottom of the conduction band remains populated by states originating from Mo and S atoms of the MoS$_2$ layers. Figure 3 shows the study on the projected density of states of the MoS$_2$/MoSe$_2$/MoS$_2$ trilayer.

In case of MoS$_2$/WSe$_2$/MoS$_2$ trilayers, similar results were obtained as seen in MoS$_2$/MoSe$_2$/MoS$_2$ trilayer. As can be seen from figure 4, the conduction band states are mostly dominated by states originating from Mo and S atoms of the MoS$_2$ layers of the trilayers. On the other hand, the valance band states appear to be dominated by states of Se and W atoms of the middle WSe$_2$ layers.

The PDOS profile of the MoS$_2$/WS$_2$/MoS$_2$ trilayer show similar characteristics as seen from the other two trilayer heterostructures. However, the difference in contributions of MoS$_2$ and WS$_2$ monolayers at the conduction band minima and valance band maxima seems to be small compared to other trilayers studied before. Although, the valance band maxima appears to be slightly dominated by states generating from middle WS$_2$ layer and conduction band minima appears to be slightly dominated by states from MoS$_2$ monolayer. Figure 5 shows the PDOS characteristics of the MoS$_2$/WS$_2$/MoS$_2$ trilayer heterostrcuture.

Electronic Properties of Trilayer Heterotructures under Strain:

We have also investigated the effect of bi-axial strain on the electronic properties of the trilayer heterostructures. First principle investigation on the electronic properties of the bilayer TMDC heterostructures reveals gradual transition from semiconducting to metallic properties as tensile strain is applied on the material (13). Applied strain also greatly modulates the nature of semiconductor bandgap. In this study, the biaxial strain is applied by stretching and compressing the lattice bi-axially and the strain is formulated as

$st = \frac{a - a_0}{a_0}$ where $a$ refers to the lattice constant of the strained lattice, $a_0$ is the lattice constant of the unstrained lattice.

Under high compressive strain, for MoS$_2$/MoSe$_2$/MoS$_2$ trialyer, the conduction band minima was found at an intermediate point ($\Delta$) between K and $\Gamma$ points in the brillouin zone, whereas the valance band maxima was found at K point. As we move from compressive to tensile strain, the conduction band valley at K point becomes more and more dominant and thereby moving the conduction band minima at K point in the brillouin zone.

On the other hand, the valance band maxima shows gradual transition from K point to $\Gamma$ point with increased tensile strain. During this gradual transition of valance band maxima and conduction band minima, for a certain range of strain, the material shows direct bandgap nature. The bandgap gradually decreases with increased tensile strain application. In our study, we found the bandgap decreasing from 0.707 eV at unstrained condition to 0.12 eV under high tensile strain for the MoS$_2$/MoSe$_2$/MoS$_2$ trilayer hetersostructure. Similar transition in conduction band minima and valance band maxima was also observed in other stacking configurations. However, the range of applied strain for which the trilayer shows direct bandgap nature varies with stacking configuration. The effect of strain in the bandstructure of MoS$_2$/MoSe$_2$/MoS$_2$ trilayer at AAA stacking configuration is shown in figure 6.

For MoS$_2$/WS$_2$/MoS$_2$ trilayer, at AAA stacking configuration, similar transition of conduction band was observed. However, for MoS$_2$/WS$_2$/MoS$_2$ heterostructure, at ABA and ACA stacking configurations, the valance band maxima was found at $\Gamma$ point for the applied strain range in this study, with the conduction band minima gradually moving from $\Delta$ point to K point in the brillouin zone. Therefore, in our range of applied strain, the MoS$_2$/WS$_2$/MoS$_2$ trialyer showed indirect bandgap nature at ABA and ACA stacking combination. The effect of strain on bandstructure of MoS$_2$/WS$_2$/MoS$_2$ trilayer under AAA stacking configuration is shown in figure 7.

On the other hand, and MoS$_2$/WSe$_2$/MoS$_2$ at AAA stacking configuration shows indirect nature only at high compressive strain. The bandgap remains direct for most of the applied range of strain. However, for ABA and ACA stacking configurations, the material structure showed indirect-direct-indirect transition and direct bandgap nature for certain range of applied strain. Figure 8 shows the change in bandstructure of MoS$_2$/WSe$_2$/MoS$_2$ trilayers at AAA stacking configuration under bi-axial strain application.

In all these cases i.e. for all these trilayers and all stacking configurations, the bandgap was found to be decreasing with applied biaxial tensile strain. The variation of bandgap with applied strain at different stacking configuration is shown in figure 9.

Study on Electron Effective Mass:

We have studied the effect of strain and stacking configurations on electron effective mass at the conduction band minima. The effective mass was calculated by fitting the *ab-initio* bandstructure at high symmetry points in the brillouin zone using parabolic approach. With increased strain, the trilayer heterostructures show decrease in electron effective mass at K point in the brillouin zone. The variation of electron effective mass at K point in the brillouin zone is shown in figure 10. The decrease in the effective mass at K point can be attributed to the gradual shift of conduction band minima toward the K point and therefore increasing the curvature of the conduction band valley at K point. Figure 11 shows the electron effective mass variation at $\Delta$ point in the brillouin zone.

## Conclusion

In this work, we have performed a first principle simulation study on trilayer $MoS_2/MX_2/MoS_2$ (M=Mo or, W; X=S or, Se) heterostructures using open source simulation package Quantum Espresso. The effect of different stacking configuration, bi-axial compressive and tensile strain on the electronic properties was observed and studied. To get better understanding of the electronic properties of these materials, projected density of states of different atoms were also studied. Of the three inserted monolayers ($WS_2$, $WSe_2$, $MoSe_2$), $WS_2$ insertion resulted in higher bandgap. Under relaxed condition, $MoS_2/MoSe_2/MoS_2$ trilayer at AAA stacking configuration shows direct bandgap nature. On the other hand, ABA and ACA stacking lead to indirect bandgap semiconductors. The study also reveals that $MoS_2/WSe_2/MoS_2$ lattice at all stacking configurations shows direct bandgap characteristics under relaxed condition. As for the $MoS_2/WS_2/MoS_2$ trilayer, the bandgap was always indirect under relaxed condition for the stacking configurations used in this study. Application of tensile strain resulted in lowering of bandgap in all the heterostructures configurations. Application of tensile strain also increases the curvature at the conduction band minima and therefore lowers the electron effective mass.

## Acknowledgments

The authors would like to thank the Department of Electrical and Electronic Engineering, Bangladesh University of Engineering and Technology for providing the facilities to carry out the simulation work.

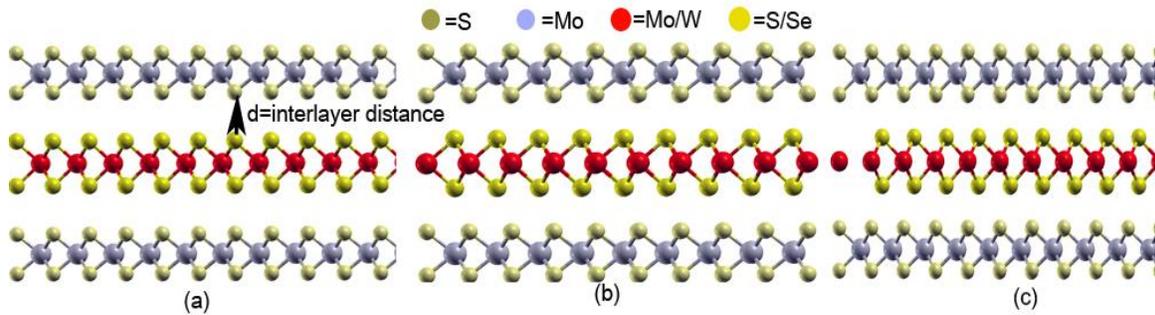

Figure 1. Stacking configurations used in this study- (a) AAA; (b) ABA; (c) ACA. Here, top and bottom layers are composed of $MoS_2$. The middle layer is composed of $MoSe_2$ or, $WS_2$ or, $WSe_2$. Different atoms of transition metals and chalcogens are shown in colors. The arrow in the figure shows interlayer distance between the two neighbouring layers.

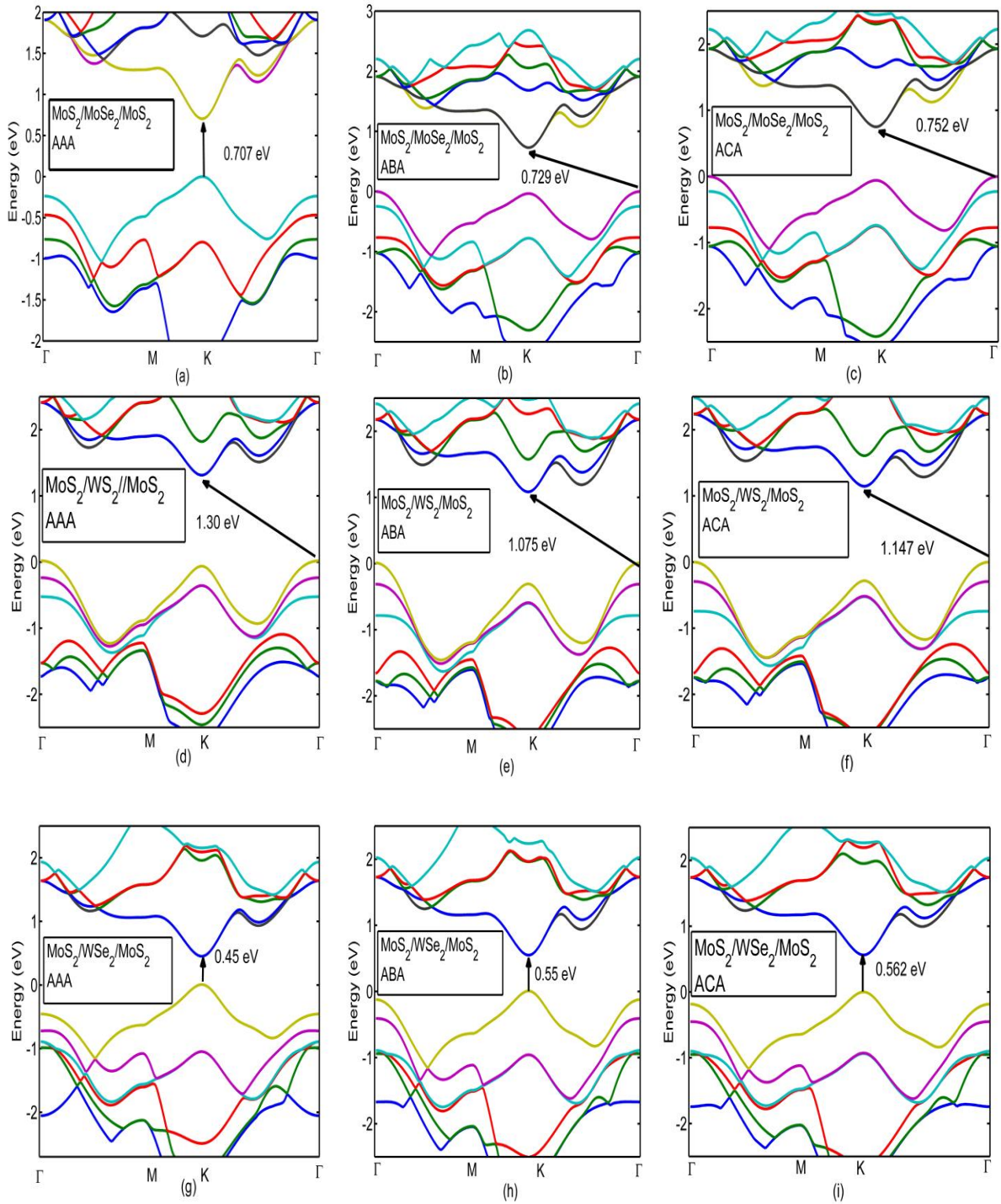

Figure 2. Extracted bandstructure for (a)-(c) $MoS_2/MoSe_2/MoS_2$; (d)-(f) $MoS_2/WS_2/MoS_2$; (g)-(i) $MoS_2/WSe_2/MoS_2$ trilayer heterostructures at different stacking configurations. As seen from the figures, inserting a $WS_2$ monolayer gives higher bandgap compared to the other two trilayer heterostructures. This higher bandgap phenomenon is also reported for $MoS_2/WS_2$ bilayer heterostrcuture. In all stacking configurations of $MoS_2/WS_2/MoS_2$ trilayer, the bandgap appears to be indirect under relaxed condition.

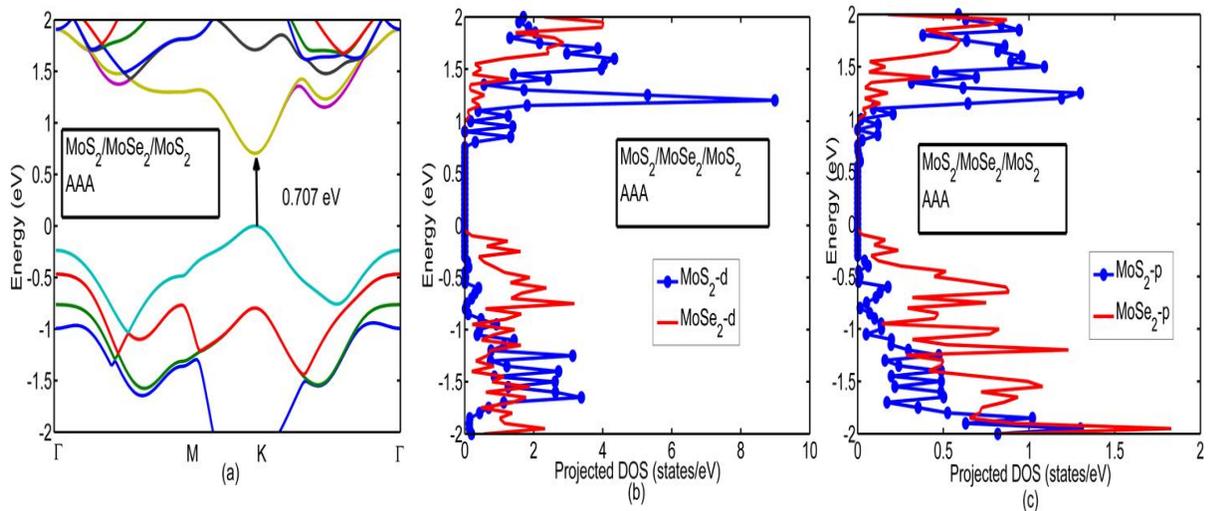

Figure 3. (a) Bandstructure of the $MoS_2/MoSe_2/MoS_2$ trilayer under AAA stacking. (b) PDOS characteristics of the trilayer due to d-electrons of the transition metal atoms. (c) PDOS characteristics of the trilayer due to p- electrons of the chalcogen atoms. As seen from the PDOS, the valance band maixma is dominated by states of middle $MoSe_2$ and conduction band states appear to be dominated by $MoS_2$ monolayer states.

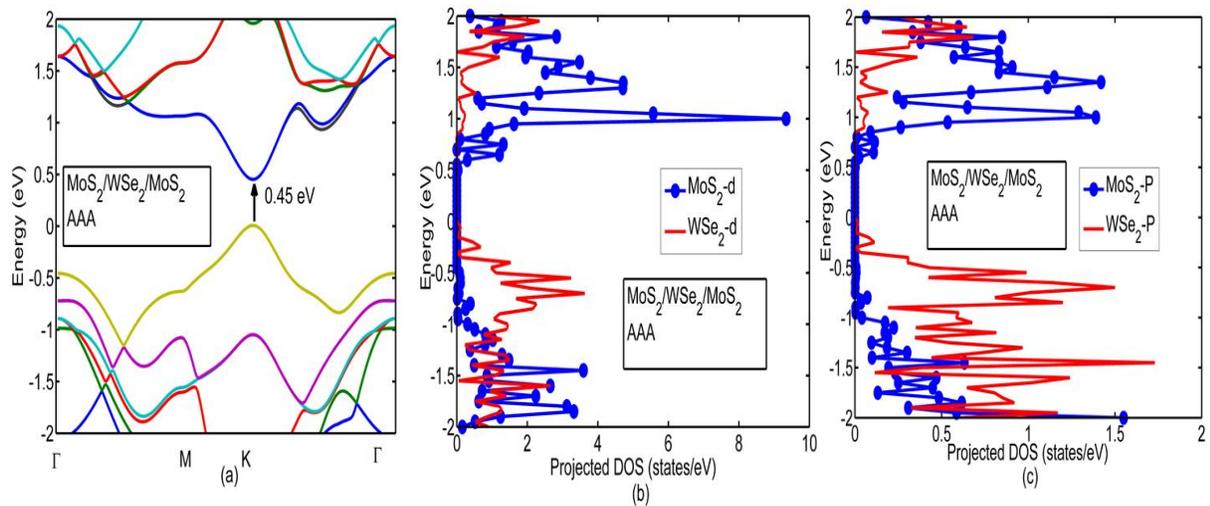

Figure 4. (a) Bandstructure of the $MoS_2/WSe_2/MoS_2$ trilayer under AAA stacking. (b) PDOS characteristics of the trilayer due to d-electrons of the transition metal atoms. (c) PDOS characteristics of the trilayer due to p-electrons of the chalcogen atoms. Like $MoS_2/MoSe_2/MoS_2$ trilayer, the conduction band appears to be dominated by states from $MoS_2$ monolayer. The valance band seems to be owed to middle $WSe_2$ monolayer.

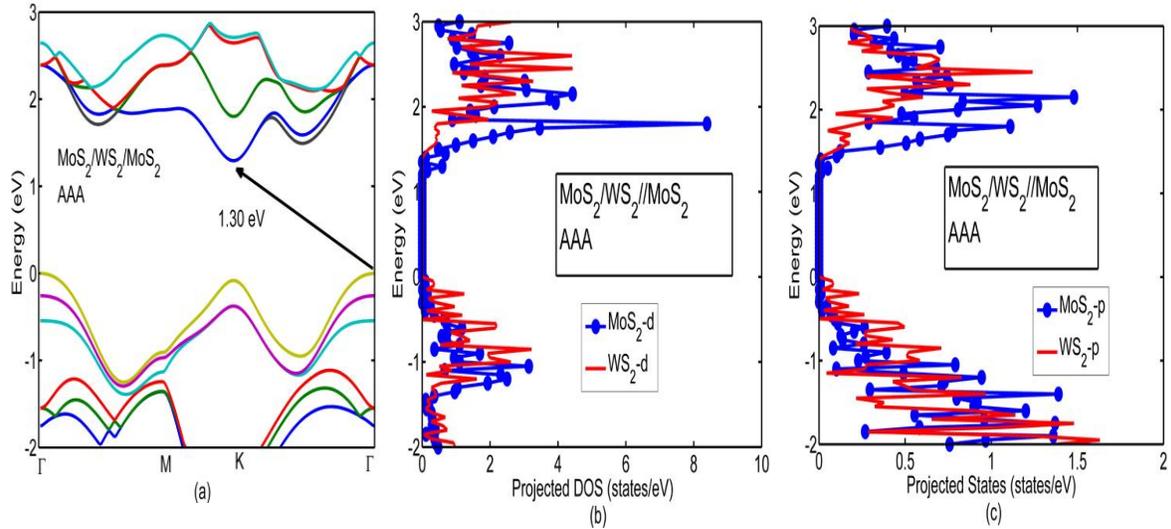

Figure 5. (a) Bandstructure of the $MoS_2/WS_2/MoS_2$ trilayer under AAA stacking. (b) PDOS characteristics of the trilayer due to d-electrons of the transition metal atoms. (c) PDOS characteristics of the trilayer due to p-electrons of the chalcogen atoms. In this case, the conduction band states appear to be slightly more owed to $MoS_2$ monolayer and middle $WS_2$ monolayer contributes to the valance band states.

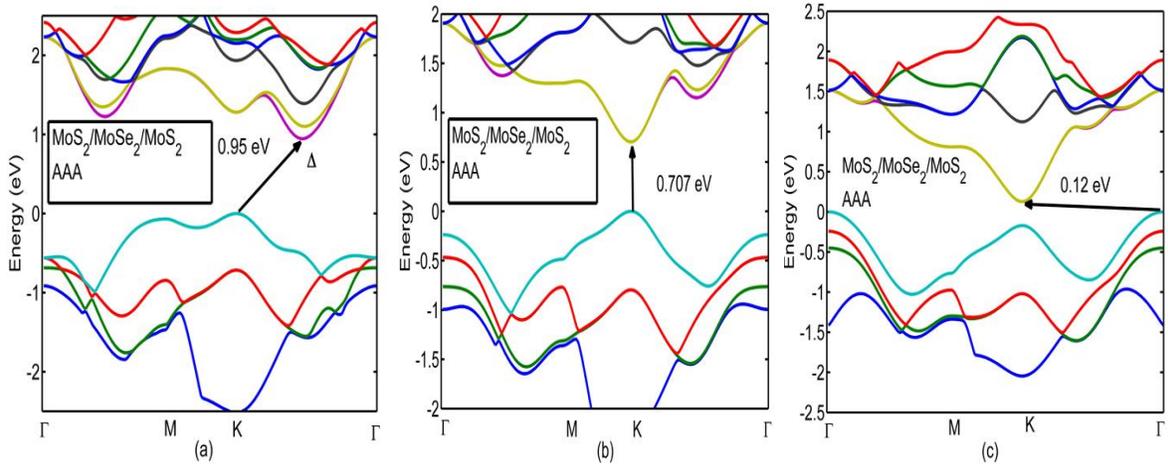

Figure 6. Modulation of bandstructure of $MoS_2/MoSe_2/MoS_2$ trilayer heterostrcture at AAA stacking configuration under bi-axial strain: (a) compressive strain; (b) unstrained lattice; (c) tensile strain. Under compressive strain, the valance band maxima is found at K point and conduction band minima is found at $\Delta$ point. As the strain approaches tensile region, the valance band shifts to the $\Gamma$ point and the conduction band moves to the K point. For a specific region of applied strain, the trilayer shows direct bandgap nature.

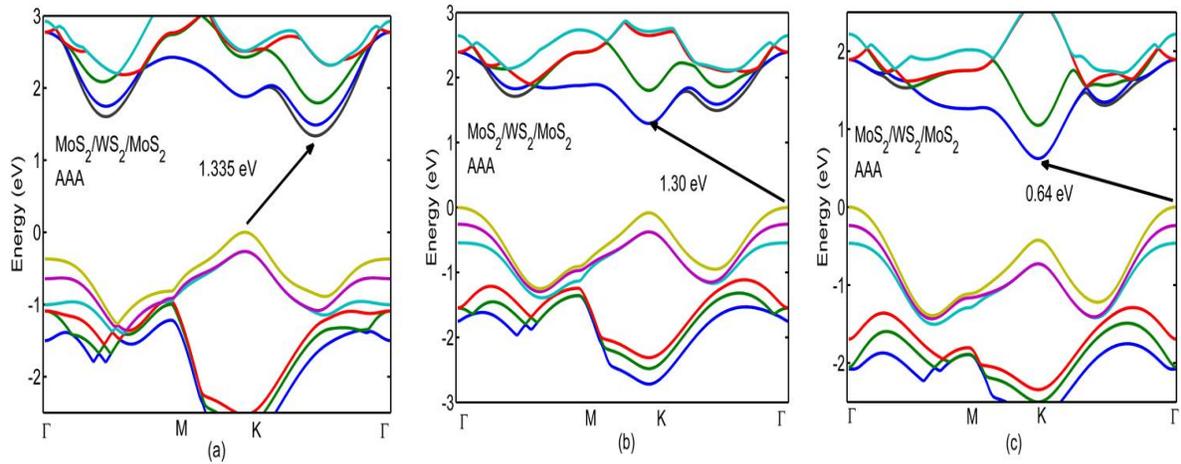

Figure 7. Modulation of bandstructure of MoS$_2$/WS$_2$/MoS$_2$ trilayer heterostrcture at AAA stacking configuration under bi-axial strain: (a) compressive strain; (b) unstrained lattice; (c) tensile strain. With increased tensile strain, the conduction band minima shifts to K point in the brillouin zone and valance band shifts from K point to $\Gamma$ point.

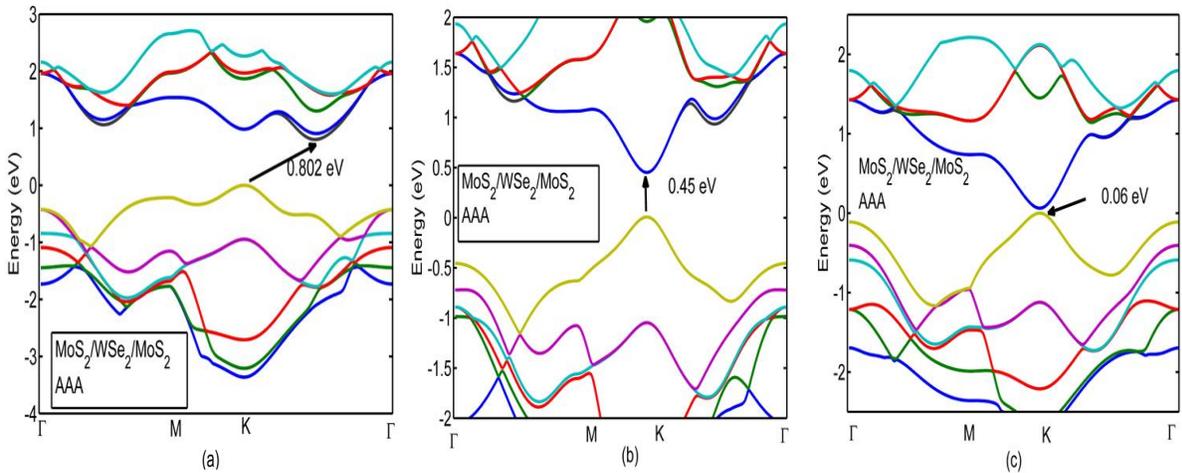

Figure 8. Modulation of bandstructure of MoS$_2$/WS$_2$/MoS$_2$ trilayer heterostrcture at AAA stacking configuration under bi-axial strain: (a) compressive strain; (b) unstrained lattice; (c) tensile strain. As seen from the figure, the valance band maxima is found at K points and the conduction band minima shifts from $\Delta$ point to K point in the brillouin zone.

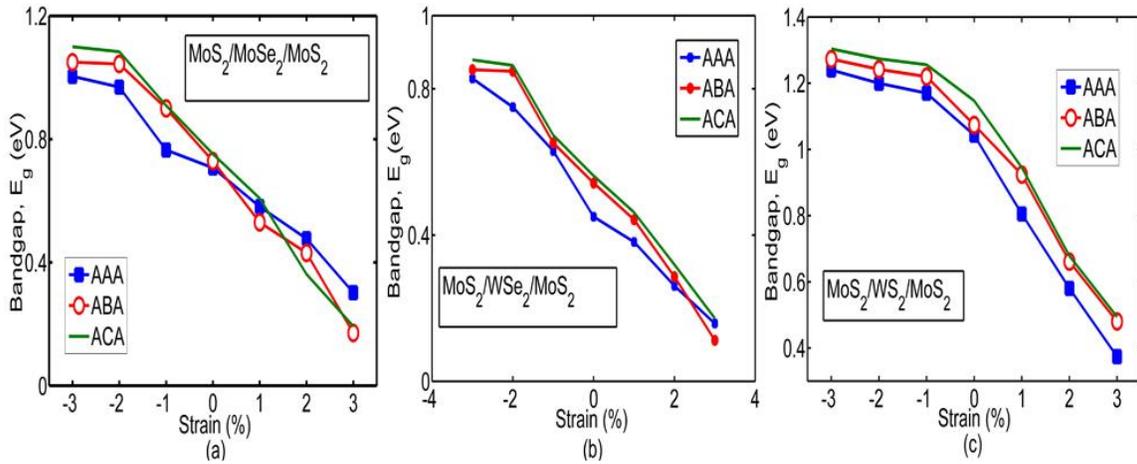

Figure 9. Variation of bandgap at different stacking configuration of (a) $MoS_2/MoSe_2/MoS_2$; (b) $MoS_2/WSe_2/MoS_2$; (c) $MoS_2/WS_2/MoS_2$ trilayer heterostructures. As can be seen, with increasing strain the heterostructures shows gradual lowering in bandgap

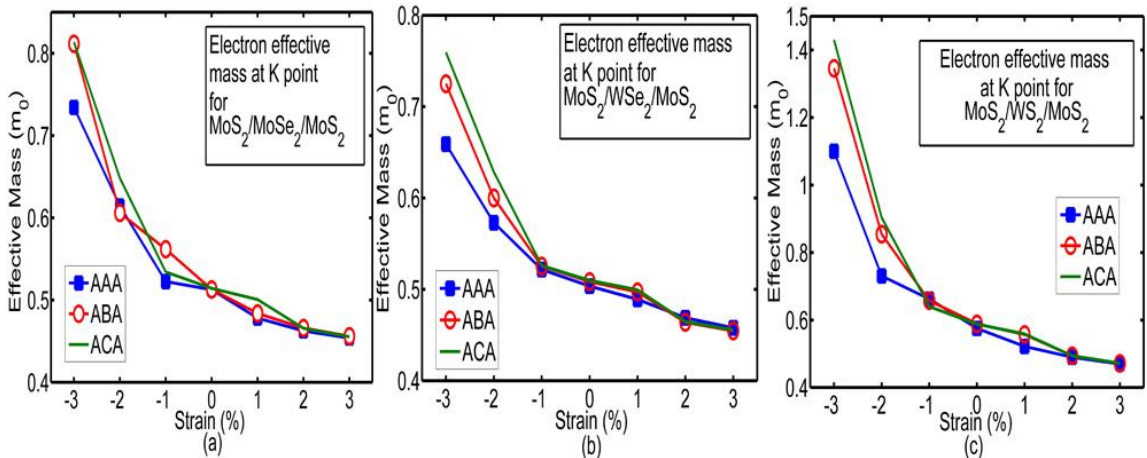

Figure 10. Electron effective mass at K point in the brilloiun zone for different stacking configuration of (a) $MoS_2/MoSe_2/MoS_2$; (b) $MoS_2/WSe_2/MoS_2$; (c) $MoS_2/WS_2/MoS_2$ trilayer heterostructures. As can be seen, the effective mass at K point decreases with increased tensile strain.

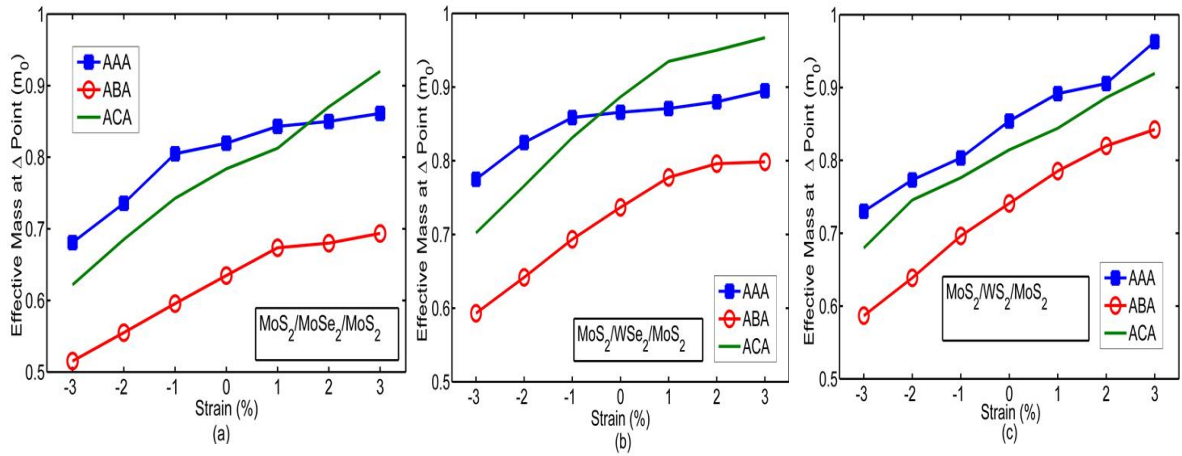

Figure 11. Electron effective mass at ∆ point in the brilloiun zone for different stacking configuration of (a) $MoS_2/MoSe_2/MoS_2$; (b) $MoS_2/WSe_2/MoS_2$; (c) $MoS_2/WS_2/MoS_2$ trilayer heterostructures. As can be seen, effective mass at ∆ point increases with strain application which indicates the shift of conduction band minima towards K point in the brillouin zone.